\documentclass[twoside,11pt]{article}

\usepackage{jmlr2e}
\usepackage[utf8]{inputenc}
\usepackage{tabularx}
\usepackage{booktabs}
\usepackage{xcolor}
\usepackage{url}
\usepackage[hidelinks]{hyperref}
\usepackage{adjustbox}
\usepackage{tikz}

\jmlrheading{0}{2015}{0-00}{11/15}{0/00}{Atılım Güneş Baydin and Barak A. Pearlmutter}

\ShortHeadings{DiffSharp: Automatic Differentiation Library}{Baydin and Pearlmutter}
\firstpageno{1}

\begin{document}

\title{DiffSharp: Automatic Differentiation Library}

\author{\name Atılım Güneş Baydin \email atilimgunes.baydin@nuim.ie\\
       \name Barak A.\ Pearlmutter \email barak@cs.nuim.ie\\
       \addr Department of Computer Science\\
       National University of Ireland Maynooth, Maynooth, Co.\ Kildare, Ireland
       \AND
       \name Jeffrey Mark Siskind \email qobi@purdue.edu\\
       \addr School of Electrical and Computer Engineering\\
       Purdue University, West Lafayette, IN 47907, USA}

\editor{}

\maketitle

\begin{abstract}
In this paper we introduce DiffSharp, an automatic differentiation (AD) library designed with machine learning in mind. AD is a family of techniques that evaluate derivatives at machine precision with only a small constant factor of overhead, by systematically applying the chain rule of calculus at the elementary operator level. DiffSharp aims to make an extensive array of AD techniques available, in convenient form, to the machine learning community. These including arbitrary nesting of forward/reverse AD operations, AD with linear algebra primitives, and a functional API that emphasizes the use of higher-order functions and composition. The library exposes this functionality through an API that provides gradients, Hessians, Jacobians, directional derivatives, and matrix-free Hessian- and Jacobian-vector products. Bearing the performance requirements of the latest machine learning techniques in mind, the underlying computations are run through a high-performance BLAS/LAPACK backend, using OpenBLAS by default.  GPU support is currently being implemented.
\end{abstract}


\section{Introduction}

Automatic differentiation (AD) is a subfield of scientific computing and specializes in calculating derivatives of functions expressed as computer programs \citep{Griewank2008}. The associativity of the chain rule of calculus leads to the two main modes of AD: the \emph{forward} (or \emph{tangent linear}) mode accumulates derivatives forward from a given independent variable, while the \emph{reverse} (or \emph{adjoint} or \emph{cotangent linear}) mode propagates derivatives backward from a given dependent variable.

AD techniques have enjoyed widespread success in computation-intensive fields such as computational fluid dynamics, atmospheric sciences, and engineering design optimization \citep{Faure2002}. The machine learning community's acquaintance with AD is mostly through the backpropagation algorithm for training feedforward neural networks, which is a special case of reverse mode AD \citep{Werbos2006,Griewank2012}.

Following a period of underutilization of general AD in machine learning, state-of-the-art machine learning frameworks increasingly provide differentiation capability in one way or another. Examples include \texttt{Theano} \citep{Bastien2012} and the recently released \texttt{torch-autograd} package for \texttt{Torch} \citep{Collobert2011}. Because the prevalent use of derivatives in machine learning is for the optimization of scalar-valued objectives, practically all of these frameworks have been restricted to reverse AD. But there is more to AD than the reverse mode.

In DiffSharp, we are introducing a framework that brings together AD with linear algebra primitives, arbitrary nesting of the forward/reverse modes, and a functional differentiation API that emphasizes the use of higher-order functions and composition. For a detailed explanation of the forward and reverse modes, the difference of AD from numerical\footnote{Finite difference approximation of derivatives, with poor performance and approximation errors.} and symbolic\footnote{Symbolic manipulation of expressions, which has problems with expression swell and control flow.} differentiation, and the use of AD in machine learning, we refer the readers to \citet{Baydin2015}.

\section{Overview of Features, Implementation, and API}

Table~\ref{TableAPI} gives an overview of the differentiation API in DiffSharp, of which extensive documentation can be found online.\footnote{\url{http://diffsharp.github.io/DiffSharp/api-overview.html}}

The library is based on AD-enabled linear algebra primitives\footnote{Scalars, vectors, and matrices, with a plan for generalizing these to tensors as in \texttt{Torch}.} that automatically compute any tangent/adjoint values associated with the forward/reverse differentiation of any forward algorithm implemented in the regular way. Programs can use the full expressivity of the language and freely make use of loops and control flow statements. The underlying computations are run on a linear algebra backend. The default backend distributed with the library uses OpenBLAS \citep{Wang2013} for BLAS/LAPACK operations, supplemented with custom parallel implementations for non-BLAS operations such as Hadamard products, elementwise function mapping, and matrix transposition.

DiffSharp supports arbitrary nesting of forward/reverse AD instantiations using tagging \citep{Siskind2008b,Pearlmutter2008} to avoid a class of bugs known as \emph{perturbation confusion} \citep{Siskind2005}. Beyond the succinctness it provides for implementing compositional models, this capability is useful for hyperparameter optimization, as it can provide exact ``hypergradients'' of the validation loss with respect to hyperparameters of training, and allow ``gradient-based optimization of gradient-based optimization'' \citep{Maclaurin2015}.

\begin{table}[t]
  \centering
  \scriptsize
  \renewcommand{\arraystretch}{0.9}
  \setlength{\tabcolsep}{1mm}
  \caption{The differentiation API for $\mathbb{R} \to \mathbb{R}$, $\mathbb{R}^n \to \mathbb{R}$, and $\mathbb{R}^n \to \mathbb{R}^m$ functions provided by the AD, numerical, and symbolic differentiation modules. X: exact; A: approximate; F: forward AD; R: reverse AD; F-R: reverse-on-forward AD; R-F: forward-on-reverse AD; F/R: forward AD if $n \le m$, reverse AD if $n > m$.}
  \label{TableAPI}
  \begin{tabularx}{\columnwidth}{@{}lllXlll@{}}
    \toprule
  	& Op. & Value & Type signature & AD & Num. & Sym.\\
    \midrule
    $f:\mathbb{R} \to \mathbb{R}$ & \texttt{diff} & $f'$ & $\color{red}{(\mathbb{R} \to \mathbb{R}) \to \mathbb{R}} \to \color{blue}{\mathbb{R}}$ & X, F & A & X\\
    &\texttt{diff'} & $(f, f')$ & $\color{red}{(\mathbb{R} \to \mathbb{R}) \to \mathbb{R}} \to \color{blue}{(\mathbb{R} \times \mathbb{R})}$ & X, F & A & X\\
    &\texttt{diff2} & $f''$ & $\color{red}{(\mathbb{R} \to \mathbb{R}) \to \mathbb{R}} \to \color{blue}{\mathbb{R}}$ & X, F & A & X\\
    &\texttt{diff2'} & $(f, f'')$ & $\color{red}{(\mathbb{R} \to \mathbb{R}) \to \mathbb{R}} \to \color{blue}{(\mathbb{R} \times \mathbb{R})}$ & X, F & A & X\\
    &\texttt{diff2''} & $(f, f', f'')$ & $\color{red}{(\mathbb{R} \to \mathbb{R}) \to \mathbb{R}} \to \color{blue}{(\mathbb{R} \times \mathbb{R} \times \mathbb{R})}$ & X, F & A & X\\
    &\texttt{diffn} & $f^{(n)}$& $\color{red}{\mathbb{N} \to (\mathbb{R} \to \mathbb{R}) \to \mathbb{R}} \to \color{blue}{\mathbb{R}}$ & X, F & & X\\
    &\texttt{diffn'} & $(f, f^{(n)})$& $\color{red}{\mathbb{N} \to (\mathbb{R} \to \mathbb{R}) \to \mathbb{R}} \to \color{blue}{(\mathbb{R} \times \mathbb{R})}$ & X, F & & X\\
    \midrule
    $f:\mathbb{R}^n \to \mathbb{R}$ & \texttt{grad} & $\nabla f$ & $\color{red}{(\mathbb{R}^n \to \mathbb{R}) \to \mathbb{R}^n} \to \color{blue}{\mathbb{R}^n}$ & X, R & A & X\\
    &\texttt{grad'} & $(f, \nabla f)$ & $\color{red}{(\mathbb{R}^n \to \mathbb{R}) \to \mathbb{R}^n} \to \color{blue}{(\mathbb{R} \times \mathbb{R}^n)}$ & X, R & A & X\\
    &\texttt{gradv} & $\nabla f \cdot \mathbf{v}$& $\color{red}{(\mathbb{R}^n \to \mathbb{R}) \to \mathbb{R}^n \to \mathbb{R}^n} \to \color{blue}{\mathbb{R}}$ & X, F & A \\
    &\texttt{gradv'} & $(f, \nabla f \cdot \mathbf{v})$ & $\color{red}{(\mathbb{R}^n \to \mathbb{R}) \to \mathbb{R}^n \to \mathbb{R}^n} \to \color{blue}{(\mathbb{R} \times \mathbb{R})}$ & X, F & A\\
    &\texttt{hessian} & $\mathbf{H}_f$& $\color{red}{(\mathbb{R}^n \to \mathbb{R}) \to \mathbb{R}^n} \to \color{blue}{\mathbb{R}^{n \times n}}$ & X, R-F & A & X\\
    &\texttt{hessian'} & $(f, \mathbf{H}_f)$ & $\color{red}{(\mathbb{R}^n \to \mathbb{R}) \to \mathbb{R}^n} \to \color{blue}{(\mathbb{R} \times \mathbb{R}^{n \times n})}$ & X, R-F & A & X\\
    &\texttt{hessianv} & $\mathbf{H}_f \mathbf{v}$ & $\color{red}{(\mathbb{R}^n \to \mathbb{R}) \to \mathbb{R}^n \to \mathbb{R}^n} \to \color{blue}{\mathbb{R}^n}$ & X, F-R & A\\
    &\texttt{hessianv'} & $(f, \mathbf{H}_f \mathbf{v})$& $\color{red}{(\mathbb{R}^n \to \mathbb{R}) \to \mathbb{R}^n \to \mathbb{R}^n} \to \color{blue}{(\mathbb{R} \times \mathbb{R}^n)}$ & X, F-R & A\\
    &\texttt{gradhessian} & $(\nabla f, \mathbf{H}_f)$ & $\color{red}{(\mathbb{R}^n \to \mathbb{R}) \to \mathbb{R}^n} \to \color{blue}{(\mathbb{R}^n \times \mathbb{R}^{n \times n})}$ & X, R-F & A & X\\
    &\texttt{gradhessian'} & $(f, \nabla f, \mathbf{H}_f)$ & $\color{red}{(\mathbb{R}^n \to \mathbb{R}) \to \mathbb{R}^n} \to \color{blue}{(\mathbb{R} \times \mathbb{R}^n \times \mathbb{R}^{n \times n})}$ & X, R-F & A & X\\
    &\texttt{gradhessianv} & $(\nabla f \cdot \mathbf{v}, \mathbf{H}_f \mathbf{v})$ & $\color{red}{(\mathbb{R}^n \to \mathbb{R}) \to \mathbb{R}^n \to \mathbb{R}^n} \to \color{blue}{(\mathbb{R} \times \mathbb{R}^n)}$ & X, F-R & A\\
    &\texttt{gradhessianv'} & $(f, \nabla f \cdot \mathbf{v}, \mathbf{H}_f \mathbf{v})$ & $\color{red}{(\mathbb{R}^n \to \mathbb{R}) \to \mathbb{R}^n \to \mathbb{R}^n} \to \color{blue}{(\mathbb{R} \times \mathbb{R} \times \mathbb{R}^n)}$ & X, F-R & A\\
    &\texttt{laplacian} & $\mathrm{tr}(\mathbf{H}_f)$ & $\color{red}{(\mathbb{R}^n \to \mathbb{R}) \to \mathbb{R}^n} \to \color{blue}{\mathbb{R}}$ & X, R-F& A & X\\
    &\texttt{laplacian'} & $(f, \mathrm{tr}(\mathbf{H}_f))$& $\color{red}{(\mathbb{R}^n \to \mathbb{R}) \to \mathbb{R}^n} \to \color{blue}{(\mathbb{R} \times \mathbb{R})}$ & X, R-F & A & X\\
    \midrule
    $\mathbf{f}:\mathbb{R}^n \to \mathbb{R}^m$ & \texttt{jacobian} & $\mathbf{J}_\mathbf{f}$ & $\color{red}{(\mathbb{R}^n \to \mathbb{R}^m) \to \mathbb{R}^n} \to \color{blue}{\mathbb{R}^{m \times n}}$ & X, F/R & A & X\\
    &\texttt{jacobian'} & $(\mathbf{f}, \mathbf{J}_\mathbf{f})$ & $\color{red}{(\mathbb{R}^n \to \mathbb{R}^m) \to \mathbb{R}^n} \to \color{blue}{(\mathbb{R}^m \times \mathbb{R}^{m \times n})}$ & X, F/R & A & X\\
    &\texttt{jacobianv} & $\mathbf{J}_\mathbf{f} \mathbf{v}$ & $\color{red}{(\mathbb{R}^n \to \mathbb{R}^m) \to \mathbb{R}^n \to \mathbb{R}^n} \to \color{blue}{\mathbb{R}^m}$ & X, F & A\\
    &\texttt{jacobianv'} & $(\mathbf{f}, \mathbf{J}_\mathbf{f} \mathbf{v})$& $\color{red}{(\mathbb{R}^n \to \mathbb{R}^m) \to \mathbb{R}^n \to \mathbb{R}^n} \to \color{blue}{(\mathbb{R}^m \times \mathbb{R}^m)}$ & X, F & A\\
    &\texttt{jacobianT} & $\mathbf{J}_{\mathbf{f}}^T$ & $\color{red}{(\mathbb{R}^n \to \mathbb{R}^m) \to \mathbb{R}^n} \to \color{blue}{\mathbb{R}^{n \times m}}$ & X, F/R & A & X\\
    &\texttt{jacobianT'} & $(\mathbf{f}, \mathbf{J}_{\mathbf{f}}^T )$ & $\color{red}{(\mathbb{R}^n \to \mathbb{R}^m) \to \mathbb{R}^n} \to \color{blue}{(\mathbb{R}^m \times \mathbb{R}^{n \times m})}$ & X, F/R & A & X\\
    &\texttt{jacobianTv} & $\mathbf{J}_{\mathbf{f}}^T \mathbf{v}$ & $\color{red}{(\mathbb{R}^n \to \mathbb{R}^m) \to \mathbb{R}^n \to \mathbb{R}^m} \to \color{blue}{\mathbb{R}^n}$ & X, R\\
    &\texttt{jacobianTv'} & $(\mathbf{f}, \mathbf{J}_{\mathbf{f}}^T \mathbf{v})$ & $\color{red}{(\mathbb{R}^n \to \mathbb{R}^m) \to \mathbb{R}^n \to \mathbb{R}^m} \to \color{blue}{(\mathbb{R}^m \times \mathbb{R}^n)}$ & X, R\\
    &\texttt{jacobianTv''} & $(\mathbf{f}, \mathbf{J}_{\mathbf{f}}^T (\cdot))$ & $\color{red}{(\mathbb{R}^n \to \mathbb{R}^m) \to \mathbb{R}^n} \to \color{blue}{(\mathbb{R}^m \times (\mathbb{R}^m \to \mathbb{R}^n))}$ & X, R\\
    &\texttt{curl} & $\nabla \times \mathbf{f}$ & $\color{red}{(\mathbb{R}^3 \to \mathbb{R}^3) \to \mathbb{R}^3} \to \color{blue}{\mathbb{R}^3}$ & X, F & A & X\\
    &\texttt{curl'} & $(\mathbf{f}, \nabla \times \mathbf{f})$ & $\color{red}{(\mathbb{R}^3 \to \mathbb{R}^3) \to \mathbb{R}^3} \to \color{blue}{(\mathbb{R}^3 \times \mathbb{R}^3)}$ & X, F & A & X\\
    &\texttt{div} & $\nabla \cdot \mathbf{f}$ & $\color{red}{(\mathbb{R}^n \to \mathbb{R}^n) \to \mathbb{R}^n} \to \color{blue}{\mathbb{R}}$ & X, F & A & X\\
    &\texttt{div'} & $(\mathbf{f}, \nabla \cdot \mathbf{f})$ & $\color{red}{(\mathbb{R}^n \to \mathbb{R}^n) \to \mathbb{R}^n} \to \color{blue}{(\mathbb{R}^n \times \mathbb{R})}$ & X, F & A & X\\
    &\texttt{curldiv} & $(\nabla \times \mathbf{f}, \nabla \cdot \mathbf{f})$ & $\color{red}{(\mathbb{R}^3 \to \mathbb{R}^3) \to \mathbb{R}^3} \to \color{blue}{(\mathbb{R}^3 \times \mathbb{R})}$ & X, F & A & X\\
    &\texttt{curldiv'} & $(\mathbf{f}, \nabla \times \mathbf{f}, \nabla \cdot \mathbf{f})$ & $\color{red}{(\mathbb{R}^3 \to \mathbb{R}^3) \to \mathbb{R}^3} \to \color{blue}{(\mathbb{R}^3 \times \mathbb{R}^3 \times \mathbb{R})}$ & X, F & A & X\\
    \bottomrule
  \end{tabularx}
\end{table}

Given our focus on higher-order functions and composition of differentiation operations, we implement the library in the F\# language\footnote{\url{http://fsharp.org/}}, an open source, cross-platform functional language with origins in ML. In addition to being a modern functional language, F\# has ``non-pure'' constructs for imperative programming and reflection, which give us more freedom for implementing features such as transformation-based AD, compared with, for example, purely functional Haskell \citep{Siskind2008b}.

Running on the .NET framework, DiffSharp can be used with F\#, C\#, and the other CLI languages. The .NET framework is now open source and cross-platform, like the .NET Core project\footnote{\url{http://dotnet.github.io/}}, which supports GNU/Linux, Mac OS X, and Microsoft Windows. We test each release of our code on Debian GNU/Linux and Microsoft Windows 10.

\section{Using the Library: Examples}

We are curating a growing collection of code and tutorials that use AD for machine learning applications.\footnote{\url{http://diffsharp.github.io/DiffSharp/}} Examples include gradient-based optimization algorithms, clustering, Hamiltonian Markov Chain Monte Carlo, and various neural network architectures.

\section {Benchmarks}

\begin{figure}
  \centering
  \trimbox{0cm 3mm}{\resizebox{0.5\textwidth}{!}{\normalsize\input{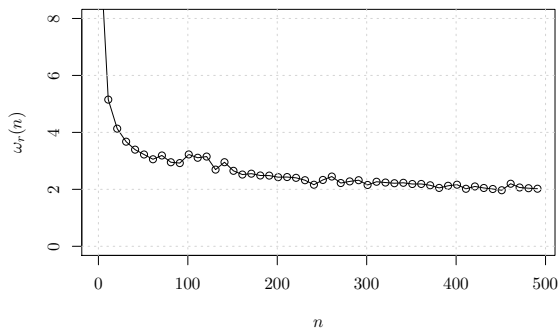}}}
  \caption{Reverse AD overhead $\omega_r(n)$ for \texttt{grad} as a function of the number of independent variables $n$ for the Helmholtz energy function.}
  \label{FigureHelmholtz}
\end{figure}

A major advantage of AD is its bounded overhead when calculating derivatives. When evaluating the gradient of a scalar-valued function with the reverse mode, the operation count of the gradient is guaranteed to be only a small constant multiple, $\omega_r$, of that of the original forward function.\footnote{In general, for a function $\mathbf{f}:\mathbb{R}^n \to \mathbb{R}^m$, if we denote the operation count to evaluate the original function $\textrm{ops}(\mathbf{f})$, we need $n \;\omega_f \;\textrm{ops}(\mathbf{f})$ to evaluate the full Jacobian $\mathbf{J} \in \mathbb{R}^{m \times n}$ with forward AD and $m \;\omega_r \;\textrm{ops}(\mathbf{f})$ with reverse AD.} Typically, the constant $\omega_r \leq 3$ and this is known as the \emph{cheap gradient principle} \citep{Griewank2008,Griewank2012}. We provide benchmarking code\footnote{\url{http://diffsharp.github.io/DiffSharp/examples-helmholtzenergyfunction.html}} for estimating $\omega_r$ as the ratio of evaluation times, for the Helmholtz energy function used in the AD literature for this purpose (Figure~\ref{FigureHelmholtz}, where $\omega_r \to 2$ as $n \to \infty$).

Similar to $\omega_r$ above for \texttt{grad}, we provide a command line benchmarking tool for reporting the overhead factors for the full array of operations in the API.\footnote{\url{http://diffsharp.github.io/DiffSharp/benchmarks.html}}

\section{Roadmap and Conclusions}

AD tools can be implemented in two main ways: source transformation and operator overloading. DiffSharp currently uses the latter with custom linear algebra primitives. An important feature we are working on is to use the ``code quotations'' meta-programming facility\footnote{Permitting the capture of type-checked expressions and effectively allowing compiler extensions.} in F\# \citep{Syme2006} for implementing a transformation-based AD where the resolution of nesting will be moved from runtime to compile time, resulting in significant performance gains and simplified user code.

We are also working on a GPU backend as an alternative to the default OpenBLAS backend, based on CUDA \citep{Nickolls2008}. Lastly, we are planning to implement advanced techniques for exploiting sparsity in linear algebra operations, using graph coloring and matrix compression techniques already developed in the AD literature \citep{Varnik2011,Walther2012}.

\acks{This work was supported in part by Science Foundation Ireland grant 09/IN.1/I2637 and US Army Research Laboratory Cooperative Agreement Number W911NF-10-2-0060.}

\setlength{\bibsep}{0.75ex}
\bibliography{baydin15a}

\end{document}